\documentstyle[A4,11pt]{article}

\def\half{\frac{1}{2}}
\def\thfs{\frac{3}{2}}
\def\tr{\mbox{\rm tr}}
\def\sign{\mbox{\rm sign}}

\begin{document}

\bigskip
\begin{center}
{\Large\bf
Shift Operator for Nonabelian Lattice Current
Algebra}
\\
\bigskip
{\large\bf
     L. Faddeev$^{1,}$\footnote[2]
     {Supported by the
 Russian Academy of Sciences and the Academy of Finland}
     and A. Yu. Volkov$^{2,}$\footnote[3]
     {On leave of absence from
Saint Petersburg Branch of the Steklov Mathematical Institute,\\
Fontanka 27, Saint Petersburg 191011, Russia}}
\end{center}
\bigskip
\begin{quote}
$^1$ {\em Saint Petersburg Branch of the Steklov Mathematical Institute\\
     Fontanka 27, Saint Petersburg 191011, Russia\\{\em and}
     Research Institute for Theoretical Physics\\
     P. O. Box 9 (Siltavuorenpenger 20C)\\
     SF-00014 University of Helsinki, Finland\\
%$^2$ Physique - Math\'{e}matique, Universit\'{e} de Montpellier II\\
%   Pl. E. Bataillon, Case 50, 34095 Montpellier C\'{e}dex 05, France}
$^2$ Laboratoire de Physique Th\'{e}orique ENSLAPP ENSLyon\\
     46 All\'{e}e d'Italie, 69007 Lyon, France}

\bigskip
{\bf Abstract.} The shift operator for a quantum lattice current
algebra associated with $sl$(2) is produced in the form of product
of local factors. This gives a natural deformation of the Sugawara
construction for discrete space-time.
\end{quote}

\section*{Introduction}

The Current Algebra provides the chiral dynamical variables for
a generic conformal field theory model called WZNW model.
Its lattice analogue, due to Semenov-Tian-Shansky,
proved useful for elucidating the quantum group structure
in this model [AFSV].
In the subsequent papers [AFS, FG, BC, AFFS] some general
properties of this algebra and its representations were discussed.
However, these considerations covered kinematical aspects of the
lattice model while such a basic dynamical object as the
hamiltonian density remained unavailable.
Here we address this problem making use of our experience in a
simpler abelian case [FV93, V95]. Following the general
philosophy worked out in these papers we construct a spatial
translation operator $W$ which simultaneously generates the
temporal shift. We find $W$ to be a product of local factors
over the lattice. This may be regarded as a multiplicative 
analogue of the Sugawara construction.

For simplicity we confine ourselves to the simplest case
of the $sl$(2) algebra. In Section 1 we recall the basic facts
about the current algebra in its classical continuous form.
Then we embed the Sugawara hamiltonian into the hierarchies of
conservation laws of two major integrable models which are
mKdV and NLS equations [FT]. To make a smoother transition to
the quantum case we present in Section 3 the
classical lattice deformation of the current algebra.
In particular, we produce relevant integrable hierarchies.
The quantum case is treated in Section 4. 

\section{Classical Model}

The generators $j^a(x)$ of current algebra are associated with
a given simple Lie algebra $g$ with index $a$ labeling
the linear basis in $g$ and the variable $x$ running through the
unit circle. Let $f^{ab}_c$ and $K^{ab}$ be the
structure constants and the Killing tensor of $g$. The defining
Poisson bracket is
$$ \{j^a(x),j^b(y)\}=\gamma f^{ab}_c j^c (x) \delta(x-y)
        +\gamma K^{ab}\delta'(x-y)   \;\;.                     $$
The real `coupling constant' $\gamma$ is irrelevant in classical
case but comes into play under quantization.

The hamiltonian
$$ H=\frac{1}{2\gamma}\int_0^{2\pi}K_{ab}j^a(x)j^b(x)dx    $$
leads to a free equation of motion
$$  \partial_t j^a(x)=\{H,j^a(x)\}=\partial_x j^a(x)     $$
which reflects the conformal invariance in hyperbolic language.
The hamiltonian density
$$ T(x)=K_{ab}j^a(x)j^b(x)   $$
is quadratic in the generators and is often referred to as the
Sugawara construction. In this paper we shall consider $g$ to be
a real form $sl(2,R)$ of the algebra $sl(2)$.

Thus, $a$ takes the values $3,+,-$ and all the functions $j^a$ are
real. The Poisson bracket is given by
$$ \{j^3(x),j^3(y)\}=\gamma\delta'(x-y)    $$
$$ \{j^3(x),j^{\pm}(y)\}=\pm\gamma j^{\pm}\delta(x-y) $$
$$ \{j^{+}(x),j^{-}(y)\}
   =2\gamma(j^{3}\delta(x-y)+\delta'(x-y)) $$
$$ \{j^{\pm}(x),j^{\pm}(y)\}=0  $$
and
$$  T=(j^3)^2+j^{+}j^{-}.   $$
It is also useful to combine the currents into a 2 by 2 matrix
$$  J=\left(\begin{array}{cc}
    j^3&j^{-}\\
    j^{+}&-j^3 \end{array}\right)=\sum j^a \sigma_a $$
and write the Poisson bracket in the form
$$  \{\stackrel{1}{J}(x),\stackrel{2}{J}(y)\}=
    \frac{\gamma}{2}[C,\stackrel{1}{J}(x)-\stackrel{2}{J}(y)]
    \delta(x-y)+\gamma C \delta'(x-y) $$
where the standard notation of the $R$-matrix formalism is
employed [FT], and $C$ is the Casimir element.

\section{Separation of Variables and Yang-Baxterization}

The above bracket and hamiltonian allow for separation of variables.
Indeed, one may put the matrix $J$ into the form
$$   J=\Omega\left(\begin{array}{cc}
     0&p\\q&0\end{array}\right)\Omega^{-1}
     +\partial_x \Omega\Omega^{-1}  $$
with a diagonal matrix $\Omega$ solving the equation
$$ \partial_x \Omega=j^3\sigma_3\Omega .   $$
The Poisson bracket for the new set of dynamical
variables $j\equiv j^3$, $p$, $q$ proves to be
$$ \{p(x),p(y)\}=-2\gamma\sign(x-y)p(x)p(y)    $$
$$ \{q(x),q(y)\}=-2\gamma\sign(x-y)q(x)q(y)    $$
$$ \{p(x),q(y)\}=2\gamma(\sign(x-y)p(x)p(y)
    +\delta'(x-y))    $$

$$ \{j(x),j(y)\}=\gamma\delta'(x-y)    $$

$$ \{j(x),p(y)\}=\{j(x),q(y)\}=0.        $$
while the hamiltonian density becomes
$$    T=j^2+pq . $$
Thus, the pair $p$, $q$ completely separates from $j$.
The $p$-$q$ bracket is known to belong to the hierarchy of Poisson
structures associated with the NLS equation while the
density $pq$ is a member (the momentum density) of the
corresponding family of
densities of local conservation laws [FT]. On the other hand,
the $j$-bracket and the density $j^2$ come from the hierarchy
of the mKdV equation. Thus, we see where $sl(2)$ current algebra
and Sugawara hamiltonian fit into the general pattern of
Soliton Theory:
$$    H_{_{\mbox{\small\rm WZNW}}}
   =P_{_{\mbox{\small\rm mKdV}}}+P_{_{\mbox{\small\rm NLS}}}.  $$
This will prove useful for our approach to
quantization.

The lattice formalism for the mKdV part, which is nothing but
the abelian current algebra, was already developed
in [FV93, V92, V95]. In this paper we perform a similar treatment
of the NLS part.

In Soliton Theory the densities of conservation laws come from
the asymptotic expansion of the trace of the monodromy matrix of
the auxiliary linear problem. For the NLS equation this problem reads
$$  \left(\partial_x+\left(\begin{array}{cc}
     0&p\\q&0\end{array}\right)+\lambda\sigma_3\right)\Psi=0.  $$
The matrix $\Omega$ being diagonal, this auxiliary problem
is gauge equivalent to
$$  \left(\partial_x+J+\lambda\sigma_3\right)\Psi=0. $$
Thus, we see that NLS part $P_{_{\mbox{\small\rm NLS}}}$ of
the Sugawara
hamiltonian is provided by `Yang-Baxterization' of the current
$$  J\leadsto J+\lambda\sigma_3.  $$
In the next Section we shall do the same on the lattice.

\section{Lattice Model}

We discretize the circle introducing the spacial
variable taking integer values running from 1 to $N$. The real
dynamical variables will be
denoted by $\alpha_{n+\half},{{\beta_{}}_{}}_n$ with integer $n$;
it is understood that
$$ \alpha_{n+N+\half}=\alpha_{n+\half}    $$
$$  \beta^{}_{n+N}=\beta^{}_{n}.          $$
One may say
that integers label vertices while half-integers stand for edges.
Or vice versa. One reason for using different notations
for dynamical variables with integer and half-integer subscripts
is mere convenience which becomes evident when the Poisson bracket
is displayed:
$$  \{\alpha_{n-\half},\alpha_{n+\half}\}
   =-2\gamma\alpha_{n-\half}\alpha_{n+\half}          $$
$$  \{\alpha_{n-\half},\beta^{}_{n}\}
   =-2\gamma\alpha_{n-\half}\beta^{}_{n}                 $$
$$   \{\beta^{}_{n},\alpha_{n+\half}\}
   =-2\gamma\beta^{}_{n}\alpha_{n+\half}          $$
$$  \{\beta^{}_{n-1},\beta^{}_{n}\}
   =2\gamma\alpha_{n-\half}\quad  .                 $$ 
All brackets not listed are zero. It is clear that every
variable
has nontrivial brackets only with the two neighbours in either
direction.

One can recognise here the so called Flaschka variables
for the Toda model. However, the hierarchy we
will deal with is different from that of the Toda equations.

To see what this lot has to do with the Current Algebra
we arrange dynamical variables in two matrices
$$ B_{2n}=
   \left(\begin{array}{cc}
   \alpha_{2n+\half}^{-\half}&0\\
   0&\alpha_{2n+\half}^{\half}\end{array}\right)
   \left(\begin{array}{cc}
   1&\beta^{}_{2n}\\&\\
   0&1\end{array}\right)  $$
$$ C_{2n-1}=
   \left(\begin{array}{cc}
   \alpha_{2n-\half}^{\half}&0\\
   0&\alpha_{2n-\half}^{-\half}\end{array}\right)
   \left(\begin{array}{cc}
   1&0\\&\\
   \beta^{}_{2n-1}&1\end{array}\right).  $$
The Poisson relations for them
$$ \{{\stackrel{1}{B}}_{2n},{\stackrel{2}{B}}_{2n}\}
   =\gamma[r_{12},{\stackrel{1}{B}}_{2n}{\stackrel{2}{B}}_{2n}] $$
$$ \{{\stackrel{1}{C}}_{2n-1},{\stackrel{2}{C}}_{2n-1}\}
   =\gamma[r_{21},{\stackrel{1}{C}}_{2n-1}{\stackrel{2}{C}}_{2n-1}] $$
$$ \{{\stackrel{1}{B}}_{2n},{\stackrel{2}{C}}_{2n-1}\}
   =\gamma{\stackrel{1}{B}}_{2n}r_{12}{\stackrel{2}{C}}_{2n-1} $$
$$ \{{\stackrel{1}{C}}_{2n+1},{\stackrel{2}{B}}_{2n}\}
   =\gamma{\stackrel{1}{C}}_{2n+1}r_{21}{\stackrel{2}{B}}_{2n} $$
employ the major ingredient
of $q$-deformations, namely the classical $r$-matrices
$$   r_{12}=\left(\begin{array}{cccc}
     \half&0&0&0\\
     0&-\half&2&0\\
     0&0&-\half&0\\
     0&0&0&\half\end{array}\right)    $$
$$   r_{21}=P_{12}r_{12}P_{12} ,   $$
where $P$ is a permutation.

The product
$$ J_n = B_{2n}C_{2n-1} $$
satisfies the Poisson brackets
$$ \{{\stackrel{1}{J}}_{n},{\stackrel{2}{J}}_{n}\}
   =\gamma(r_{12}{\stackrel{1}{J}}_{n}{\stackrel{2}{J}}_{2n}
   -{\stackrel{1}{J}}_{n}{\stackrel{2}{J}}_{2n}r_{21}) $$
$$ \{{\stackrel{1}{J}}_{n+1},{\stackrel{2}{J}}_{n}\}
   =\gamma{\stackrel{1}{J}}_{n+1}r_{21}{\stackrel{2}{J}}_{n} $$
which turn into the Current Algebra in the continuum limit
$$ J_{n}\sim I+\Delta J(x). $$
This is what usually is called the Lattice Current Algebra.
However, it is not clear whether one gains anything reducing
$B$-$C$-algebra to the $J$-one. This time we prefer to deal
with somewhat more transparent $B$-$C$-algebra but we could
do with the $J$-one instead.

To produce relevant conservation laws we introduce the
so called `transfer-matrix'
$$  t(\omega)=\tr\prod^{\leftarrow}_{n}
    \xi^{\sigma_3}B_{2n}\eta^{-\sigma_3}C_{2n-1}   $$
with `spectral parameter' $\omega$ entering in $\xi$ and $\eta$ in
such a way that
$$   \xi^2+\eta^2=2    $$
$$   \frac{\xi}{\eta}=\omega .     $$
It turns out that
\begin{itemize}
  \item[(i)] $t(\omega)$ is a Poisson commuting family:
             $$   \{t(\omega),t(\omega')\}=0 ,    $$
  \item[(ii)] in the continuous limit it turns into the
trace of monodromy matrix of
the continuous auxiliary linear problem of Section 1 provided
$$  \omega\sim 1+\Delta\lambda  , $$
  \item[(iii)] it is a power series in $\omega$
$$ t(\omega)=\sum_{-N/2}^{N/2}h_k \omega^{2k} $$
with
$$ h_{N/2}=\prod_n\left(2+
   \frac{\beta^{}_{2n+1}\beta^{}_{2n}}
   {\alpha_{2n+\half}}\right) $$
$$ h_{-N/2}=\prod_n\left(2+
   \frac{\beta^{}_{2n}\beta^{}_{2n-1}}
   {\alpha_{2n-\half}}\right). $$
\end{itemize}
(ii) is obvious, (iii) is almost so, (i) can be verified along the
guidelines of [FM]. We shall not go into further details because
the model in question actually belongs to the same hierarchy as the
Ablowitz-Ladik's model [SV]. 

In the continuous limit we have
$$  H=\log h_{N/2} + \log h_{-N/2}\sim \Delta\int_0^{2\pi}
    J^{+}J^{-}dx=\Delta P_{_{\mbox{\rm NLS}}} $$
as should be expected.

We have obtained the hamiltonian of the classical lattice model
which plays the role of the NLS part of the Sugawara construction
for the lattice current algebra. The corresponding mKdV part
can be found in [V92]. However, the equations of
motion produced by these hamiltonians are quite complicated
and turn into simple free equations only in the continuous
limit. It was realized in [FV93] that the discrete time equation
$$ J_n(t+\Delta)=J_{n+1}(t) $$
is a better option. In other words, the discretizing of
space should be accompanied by the discretizing of time. The last
equation is especially transparent in the quantum theory where
the spacial shift operator $W$ such that
$$ W^{-1}J_n W=J_{n+1} $$
is taken to define the time shift as well
$$ J_{n}(t+\Delta)= W^{-1}J_n(t) W $$
$$ J_{n}(0)=J_{n}.                 $$
We shall find this operator in the next
Section. The expression for the classical lattice hamiltonian
will prove to be a useful hint in our search.

\section{Shift Operator}

The quantum lattice current algebra inherits the
notation $\alpha$-$\beta$ for generators together with the
way they are enumerated while the Poisson relations turn into
their most natural quantum counterparts
$$ \alpha_{n+\half}\alpha_{n-\half}
   =q^2\alpha_{n-\half}\alpha_{n+\half}          $$
$$ \beta^{}_{n}\alpha_{n-\half}
   =q^2\alpha_{n-\half}\beta^{}_{n}              $$
$$ \alpha_{n+\half}\beta^{}_{n}
   =q^2\beta^{}_{n}\alpha_{n+\half}              $$
$$  [\beta^{}_{n-1},\beta^{}_{n}]
   =(q-q^{-1})\alpha_{n-\half}\quad  ,           $$ 
with the deformation parameter $q$ combining the coupling
constant $\gamma$ and the Planck constant $\hbar$ in the usual way
$$ q=e^{i\hbar\gamma}. $$
The consistency of these commutation
relations becomes more apparent as soon as one rewrites them
in $R$-matrix form
$$ R_{12}{\stackrel{1}{B}}_{2n}{\stackrel{2}{B}}_{2n}
   ={\stackrel{2}{B}}_{2n}{\stackrel{1}{B}}_{2n} R_{12}    $$
$$ R_{21}{\stackrel{1}{C}}_{2n-1}{\stackrel{2}{C}}_{2n-1}
   ={\stackrel{2}{C}}_{2n-1}{\stackrel{1}{C}}_{2n-1} R_{21}$$
$$ {\stackrel{2}{C}}_{2n-1}{\stackrel{1}{B}}_{2n}
   ={\stackrel{1}{B}}_{2n}R_{12}{\stackrel{2}{C}}_{2n-1}   $$
$$ {\stackrel{2}{B}}_{2n}{\stackrel{1}{C}}_{2n+1}
   ={\stackrel{1}{C}}_{2n+1}R_{21}{\stackrel{2}{B}}_{2n}   $$
where matrices $B$,$C$ are built of $\alpha$,$\beta$'s in
literally the same way as in the classical case of Section 2.
The $R$-matrix involved is the $sl(2)$ one
$$   R_{12}=\left(\begin{array}{cccc}
     q^{\half}&&&\\
     &q^{-\half}&q^{\half}-q^{-\frac{3}{2}}&\\
     &&q^{-\half}&\\
     &&&q^{\half}\end{array}\right)    $$
and it is needless to say that the associativity of
the $B$-$C$-algebra is due to the Yang-Baxter equation
$$ R_{12}R_{13}R_{23}=R_{23}R_{13}R_{12} $$
fulfilled by $R$.

The way variables separate in the continuous model and
the belief that integrals of local densities on the
lattice turn into products of local factors suggest that the
shift operator
$$ \alpha_{n-\half}W=W\alpha_{n+\frac{3}{2}} $$
$$ \beta^{}_{n-1}W=W\beta^{}_{n+1}  $$
should decompose into a product of two commuting factors
$$    W=UV=VU       $$
of the form
$$ U=\prod^{\leftarrow}
   \theta(q\alpha_{n+\half}^{}\alpha_{n-\half}^{-1}) $$
$$ V=\prod^{\leftarrow}\sigma(t_{n-\half}) $$
$$ t_{n-\half}=q+q^2\beta^{}_{n}\alpha^{-1}_{n-\half}
                    \beta^{}_{n-1}     .              $$
It only remains to find suitable functions $\theta$ and $\sigma$.
It is shown in Appendix B that the functions doing the job are
$$ \theta(z)=\exp\frac{(\log (-z))^2}{4\log q} $$
and
$$ \sigma(z)=\exp\frac{1}{4}\int_{-\infty}^{\infty}
   \frac{z^{i\zeta}}{\sinh(\pi\zeta)\sinh(\gamma\hbar\zeta)}
   \frac{d\zeta}{\zeta} $$
where the contour of integration rounds the singularity
at $\zeta=0$ from above.

Let us conclude with some remarks.
It must be said that although the solution looks like a
straightforward remake of the one for the $U(1)$ case [FV93] it
is not. The logics of the abelian case would rather favour another
local decomposition for the shift operator, with
functions $\theta(-z)$ and $\sigma(-z)$ instead
of  $\theta(z)$ and $\sigma(z)$. Of course, the minus in the
argument of $\sigma$ is a lot more important than it may look.
In particular, the r.h.s. of the functional equation fulfilled
by $\sigma(z)$ (see Appendix) comes directly from the density
of the classical lattice hamiltonian
$$   \frac{1}{1+t}\sim\frac{1}{2+\beta\beta/\alpha}.  $$
This correspondence principle plays a major role in the detailed study
of the classical limit which will be presented elsewhere.

As we said, one could do with the $J$-picture from the very
beginning. This would eventually lead to the following decomposition
$$ W=\prod^{\leftarrow}\sigma(t_{2n+\half})
     \theta(q\alpha_{2n+\half}^{}\alpha_{2n-\half}^{-1})
     \sigma(t_{2n-\half})     $$
for the $J$-shifting operator
$$ J_n W= W J_{n+1}. $$
It is easy to check that all entries of this decomposition
do belong to the $J$-subalgebra of the $B$-$C$-algebra. However,
it is not that apparent why the reduction $B$-$C\rightarrow J$ should
eliminate some factors in $W$ and modify some of remaining ones.
Subtle things like central elements are partly responsible for this.
Which in turn has a lot to do with another important issue which
we deliberately left aside. Indeed, we did not pay proper attention
to the periodic boundary condition. What was then meant by
ordered products over the lattice? For the answer to this question
we refer the reader to [V95].

\section*{Appendix. Full Pull}

In this paper, as well as on a few earlier
occasions [FV93, FV94, F, FV95], we relied
on certain `$q$-special-functions' satisfying functional
equations of the form
$$   \frac{f(qz)}{f(q^{-1}z)}=d(z).       $$
In particular, the functions $\theta$ and $\sigma$ of the
last Section were given
by $d(z)=z$ and $d(z)=\frac{1}{1+z}$ respectively.    
Although the issues of the solvability and the good choice
of solution are important in their own right, for practical
purposes (read formal computations) one
seldom needs anything else than just the equation
itself. Let us show how it goes in our case.

Due to the locality of commutation relations the equations
$$ \alpha_{n-\half}W=W\alpha_{n+\frac{3}{2}} $$
$$ \beta^{}_{n-1}W=W\beta^{}_{n+1}  $$
easily reduce to
$$  \alpha_{n-\half}^{}
    \theta(q\alpha_{n+\thfs}^{}\alpha_{n+\half}^{-1})
    \theta(q\alpha_{n+\half}^{}\alpha_{n-\half}^{-1})
   =\theta(q\alpha_{n+\thfs}^{}\alpha_{n+\half}^{-1})
    \theta(q\alpha_{n+\half}^{}\alpha_{n-\half}^{-1})
    \alpha_{n+\thfs}^{}.                             $$
and
$$ \beta^{}_{n-1}\sigma(t_{n+\half})
   \theta(q\alpha_{n+\half}^{}\alpha_{n-\half}^{-1})
   \sigma(t_{n-\half})=\sigma(t_{n+\half})
   \theta(q\alpha_{n+\half}^{}\alpha_{n-\half}^{-1})
   \sigma(t_{n-\half}) \beta^{}_{n+1}  .                    $$
The mere form of the functional equation suggests that any
computation ought to be a sequence of just two elementary steps:
\begin{itemize}
   \item[(i)]$w f(z)=f(z) \:w \qquad$if $wz=zw$,
   \item[(ii)]$x f(y)=f(y)\:d(qy) x
               \qquad$if $x,y$ make a Weyl pair: $xy=q^2 yx$.
\end{itemize}
Since any two of the $\alpha$'s either commute or make a Weyl pair, the
first translation comes easy:
$$  \alpha_{n-\half}^{}
    \theta(q\alpha_{n+\thfs}^{}\alpha_{n+\half}^{-1})
    \theta(q\alpha_{n+\half}^{}\alpha_{n-\half}^{-1}) $$ 
$$  =-q^{2}\theta(q\alpha_{n+\thfs}^{}\alpha_{n+\half}^{-1})
    \alpha_{n+\thfs}^{}\alpha_{n+\half}^{-1}\alpha_{n-\half}^{}
    \theta(q\alpha_{n+\half}^{}\alpha_{n-\half}^{-1}) $$
$$  =q^{4}\theta(q\alpha_{n+\thfs}^{}\alpha_{n+\half}^{-1})
    \theta(q\alpha_{n+\half}^{}\alpha_{n-\half}^{-1})
    \alpha_{n+\half}^{}\alpha_{n-\half}^{-1}
    \alpha_{n+\thfs}^{}\alpha_{n+\half}^{-1}\alpha_{n-\half}^{} $$
$$  =\theta(q\alpha_{n+\thfs}^{}\alpha_{n+\half}^{-1})
    \theta(q\alpha_{n+\half}^{}\alpha_{n-\half}^{-1})
    \alpha_{n+\thfs}^{}.                             $$

The second one is more tricky. We cannot
pull $\beta^{}_{n-1}$ through $\sigma(t_{n+\half})$ straight away
because $\beta^{}_{n-1}$ and $t_{n+\half}$ neither commute
nor make a Weyl pair. Nevertheless, we have a good supply of
operators making `good' pairs
with both $t_{n-\half}$ and $t_{n+\half}$ which, by the way,
between themselves are a $q$-oscillator\footnote{The remaining
nontrivial commutation relations governing the algebra of $t$'s,
those for the neighbours twice removed, seldom participate
in computations. Their role may be seen in taking care of the
associativity of the algebra. Anyway, their explicit form
can be found in [V92]. This algebra is sometimes referred to
as the Lattice Virasoro Algebra for in a certain continuous
limit, different from the one of the present paper, it turns into
the Virasoro algebra with a nonzero central charge.}
$$ q t_{n+\half}t_{n-\half}-q^{-1}t_{n-\half}t_{n+\half}
   =q-q^{-1}    .                                $$
Among them are:
\begin{itemize}
  \item[(i)] all the $\alpha$'s
$$ [\alpha,t]=0, $$
  \item[(ii)] the $\beta$ which is `between' them
$$ t_{n-\half}\beta^{}_{n}=q^2 \beta^{}_{n}t_{n-\half} $$
$$ \beta^{}_{n}t_{n+\half}=q^2 t_{n+\half}\beta^{}_{n},$$
  \item[(iii)] another operator $c^{}_{n}
=q(t_{n+\half}t_{n-\half}-1)$
$$ t_{n-\half}c^{}_{n}=q^2 c^{}_{n}t_{n-\half} $$
$$ c^{}_{n}t_{n+\half}=q^2 t_{n+\half}c^{}_{n} $$
which is a familiar satellite of $q$-oscillators.
\end{itemize}
So, we express $\beta^{}_{n-1}$ via `good' operators
$$ \beta^{}_{n-1}
   =qt^{-1}_{n+\half}c^{}_n\beta^{-1}_{n}\alpha_{n-\half}
   +q^{2}t^{-1}_{n+\half}\beta^{-1}_{n}\alpha_{n-\half}
   -q\beta^{-1}_{n}\alpha_{n-\half}                $$
and get
$$ \beta^{}_{n-1}\sigma(t_{n+\half})          $$
$$   =\sigma(t_{n+\half})
   \left(q t^{-1}_{n+\half}c^{}_n\beta^{-1}_{n}\alpha_{n-\half}
   +(1+q^{-1}t_{n+\half})
   (q^{2}t^{-1}_{n+\half}\beta^{-1}_{n}\alpha_{n-\half}
   -q\beta^{-1}_{n}\alpha_{n-\half})\right)   $$
$$ =\sigma(t_{n+\half})
   \left( (\beta^{-1}_{n}t_{n-\half}
   -t_{n+\half}\beta^{-1}_{n})\alpha_{n-\half})\right). $$
Similarly,
$$ \sigma(t_{n-\half})\beta^{}_{n+1}
   =\left(\alpha_{n+\half}(-\beta^{-1}_{n}t_{n-\half}
   +t_{n+\half}\beta^{-1}_{n})\right)\sigma(t_{n-\half}). $$
The rest
$$ \left( (\beta^{-1}_{n}t_{n-\half}
   -t_{n+\half}\beta^{-1}_{n})\alpha_{n-\half})\right)
   \theta(q\alpha_{n+\half}^{}\alpha_{n-\half}^{-1})    $$
$$  =\theta(q\alpha_{n+\half}^{}\alpha_{n-\half}^{-1})
  \left(\alpha_{n+\half}(-\beta^{-1}_{n}t_{n-\half}
   +t_{n+\half}\beta^{-1}_{n})\right)    $$
is a variation on the same theme.

Unfortunately, all this looks like a hat-trick rather
than conscious approach. A more systematic paper making
better use of YangBaxterization is on the authors' agenda.

\begin{quote}
{\bf Acknowledgements.}
We are grateful to
A. Alekseev and M. Semenov-Tian-Shansky for stimulating
discussions and to S. Marshakov for urging us to complete
this work.
\end{quote}
\section*{\normalsize\bf References}

\begin{itemize}
  \item[{[AFSV]}] A. Alekseev, L. Faddeev,
M. Semenov-Tian-Shansky, A. Volkov, preprint CERN-TH-5981/91.
  \item[{[AFS]}] A. Alekseev, L. Faddeev,
M. Semenov-Tian-Shansky, Comm. Math. Phys. 149 (1992) 335.
  \item[{[AFSV]}] A. Alekseev, L. Faddeev,
J. Fr\"{o}hlich, V. Shomerus, q-alg/9604017.
  \item[{[BC]}] A. Belov and K. Chaltikian,
Phys. Lett. B317 (1993) 73.
  \item[{[F]}] L. Faddeev, hep-th/9408041.
  \item[{[FT]}] L. Faddeev, L. Takhtajan, Hamiltonian
methods in the theory of solitons (Springer, Berlin, 1987).
%  \item[{[FV92]}] L. Faddeev and A. Volkov,
%Theor. Math. Phys. 92 (1992) 207.
  \item[{[FV93]}] L. Faddeev and A. Volkov, Phys. Lett. B315 (1993) 311.
  \item[{[FV94]}] L. Faddeev and A. Volkov,
Lett. Math. Phys. 32 (1994) 125.
  \item[{[FV95]}] L. Faddeev and A. Volkov, Zap. Nauchn.
Semin. PDMI 224 (1995).
  \item[{[FG]}] F. Falceto and K. Gawedzky,
J. Geom. Phys. 11 (1993) 251.
  \item[{[FM]}] L. Freidel and J. M. Maillet,
Phys. Lett. B263 (1991) 403.
  \item[{[SV]}] E. Sklyanin and A. Yu. Volkov, unpublished.
  \item[{[V92]}] A. Volkov, Phys. Lett A167 (1992) 345.
  \item[{[V95]}] A. Volkov, hep-th/9509024
\end{itemize}
\end{document}